\documentclass[twocolumn,aps,showpacs,floatfix]{revtex4}
\usepackage{graphicx}
\usepackage{bm}

\begin{document}

\title{Topology of the ground state of two interacting 
Bose-Einstein condensates}

 \author{Francesco Riboli}
  \email{riboli@lens.unifi.it}
 \author{Michele Modugno}
  \email{modugno@fi.infn.it}
 \affiliation{%
 INFM - LENS - Dipartimento di Fisica, Universit\`a di Firenze\\
 Via Nello Carrara 1, 50019 Sesto Fiorentino, Italy
 }%

\date{\today}

\begin{abstract}
We investigate the spatial patterns of the ground state 
of two interacting Bose-Einstein condensates. 
We consider the general case of two different atomic species
(with different mass and in different hyperfine states) trapped in a
magnetic potential whose eigenaxes can be tilted with respect
to the vertical direction, giving rise to a non trivial gravitational sag. 
Despite the complicated geometry, we show that within the Thomas-Fermi 
approximations and upon appropriate coordinate transformations, the 
equations for the density distributions can be put in a very simple form. 
Starting from this expressions we give explicit rules to classify
the different spatial topologies which can be produced, 
and we discuss how the behavior of the system is influenced by the 
inter-atomic scattering length. 
We also compare explicit examples with
the full numeric Gross-Pitaevskii calculation.

\end{abstract}
\pacs{03.75.Fi, 05.30.Jp}
\maketitle

\section{Introduction}
\label{sec:intro}

Bose-Einstein condensation of mixtures of different atomic species
has recently been the subject of an intensive experimental and 
theoretical research 
\cite{jila1,jila2,otago,lens,mit,ho,bigelow,timm,ao,esry,tripp,barankov}.
The first experimental realization of a system of two interacting 
Bose-Einstein condensates (BECs)
has been obtained at JILA with a double condensate of $^{87}$Rb in two 
different hyperfine states, $|F,M_F\rangle=|1,-1\rangle$
and $|2,2\rangle$ \cite{jila1}. 
This mixture was characterized by a partial overlap between the two
condensates, in presence of a gravitational ``sag'' due to the different 
magnetic moment.
Since then several other experiments have been performed
with double condensates of rubidium \cite{jila2,otago,lens} 
and with spinor condensates of sodium in optical traps \cite{mit}.

Motivated by these experiments and by
the future possibility of realizing other
binary mixtures of interacting BECs, these systems have been 
extensively studied also from
the theoretical point of view. Up to now only two particular cases have
been addressed: {\em(i)} a system of two condensates with different mass 
in cylindrically symmetric potentials arranged concentrically 
\cite{ho,bigelow,tripp,barankov}, and
{\em(ii)} including a gravitational sag, 
but for condensates with the same mass (the JILA case) \cite{esry}. 

In this paper we extend these studies by considering the very general case 
of two different atomic species, with different mass and in different 
hyperfine states, trapped in a magnetic potential whose eigenaxes can be 
tilted with respect to the direction of gravity.
We show that, despite the complicated geometry, the ground-state of the
system can be easily characterized within the Thomas-Fermi approximation valid 
for large numbers of atoms. We  provide general formulas which allow
to calculate the shape and the density distributions of the two BECs.
Our results can be a useful tool to analyze future experiments.

The paper is organized as follows.
In Section \ref{sec:model} we discuss the  equations for
the ground state of the system and show
that upon an appropriate coordinate transformation, they 
can be put in a very simple spherical form.
In Section \ref{sec:tf} we discuss the general features of the model, 
and we give an explicit algorithm to classify all the different 
topologies which can be constructed by varying the number of atoms and
the inter-atomic scattering length.
We also work out some example for the case
$^{87}$Rb and $^{41}$K, which is a promising system for the 
realization of a new binary mixture of BECs \cite{potassio,potassio_new}.
Finally we compare the results against
the numerical solution of the full Gross-Pitaevskii equations
for the system, finding a good agreement.

\section{The model}
\label{sec:model}

Let us consider a system of two Bose-Einstein condensates
with mass $m_i$ and in the hyperfine state $(F_i,M_{Fi})$, 
each containing $N_i$ atoms ($i=1,2$), 
confined in a magnetic trap.
The ground state of the system can be obtained by solving two
coupled Gross-Pitaevskii equations for the condensate wave-functions 
$\psi_i$ \cite{report}
\begin{equation}
\left[-{\hbar^2\over2m_1}\nabla^2 + U_1({\bf x})+ 
u_{11}|\psi_1|^2+u_{12}|\psi_2|^2\right]\psi_1=\mu_{1}\psi_1
\label{eq:gpe1}
\end{equation}
\begin{equation}
\left[-{\hbar^2\over2m_2}\nabla^2 + U_2({\bf x})+ 
u_{21}|\psi_1|^2+u_{22}|\psi_2|^2\right]\psi_2=\mu_{2}\psi_2
\label{eq:gpe2}
\end{equation}
with the normalization condition
\begin{equation}
\int d^3x |\psi_i|^2=N_i \,.
\label{eq:norma}
\end{equation}
The coupling constants $u_{ij}$ are given in terms of the scattering
length $a_{ij}$ by \cite{esry}
\begin{eqnarray}
u_{11}&=&{4\pi\hbar^2 a_{11}\over m_1}\\
u_{12}&=&2\pi\hbar^2 a_{12}\left({m_1+m_2\over m_1 m_2}\right)=u_{21}\\
u_{22}&=&{4\pi\hbar^2 a_{22}\over m_2}
\end{eqnarray}
where we used the fact that $a_{12}=a_{21}$.
Hereinafter we assume $a_{11}$, $a_{22} >0$.

The total potential experienced by each condensate is
the sum of the gravitational potential and of a dipole magnetic 
potential $U_B^{i}({\bf x})=\mu_B g_{Fi} M_{Fi} |B({\bf x})|$
($g_{Fi}$ is the gyro-magnetic factor of the specie $i$) which we
assume, as usual,  to be harmonic
\begin{equation}
U_B^{(i)}({\bf x})
=\mu_B g_{Fi} M_{Fi} \left(B_0 +\frac{1}{2}\sum_j K_j x^2_j\right)
\end{equation}
By defining $\bar{K}=(K_1K_2K_3)^{1/3}$, 
$\lambda_j=\sqrt{K_j/\bar{K}}$, and
\begin{eqnarray}
U_{0i}&=&\mu_B g_{Fi} M_{Fi} B_0\\
\omega_i^2&=&\mu_B g_{Fi} M_{Fi}\bar{K}/m_i
\end{eqnarray}
we can finally write $U_B$ in the standard form
\begin{equation}
U_B^{(i)}({\bf x})={1\over2}m_i\omega_i^2\sum_j \lambda^2_j x^2_j + U_{0i}
\end{equation}
For what concerns the gravitational potential,
here we consider the general case in which the vertical 
direction (the direction of gravity) 
is not aligned with any of the symmetry axis of the trap, 
but lies in the $x-z$ symmetry plane, rotated by an angle $\theta$.
We include this possibility since in the experiments the 
trap confinement is generally weaker along the horizontal direction $x$,
and therefore even a small angle can produce a large ``horizontal sag''; 
we will give explicit examples in the following section \cite{nota}. 
The total potential is
\begin{equation}
U_i({\bf x})=U_B^{(i)}({\bf x})+ m_i  g (x\sin\theta+z\cos\theta)
\end{equation}

By performing an appropriate transformation on the coordinates
the potential $U(\bf{x})$ can be put in a simpler form.
These transformations amount to
\begin{itemize}

\item[(i)]
a scaling by $\lambda_j$ (notice that the determinant of
the transformation is equal to one)
\begin{equation}
x_j\longrightarrow x'_j\equiv\lambda_j x_j
\label{eq:scaling}
\end{equation} 
in order to put $U_B(\bf{x}')$ in a spherically symmetric form;

\item[(ii)]
a rotation of an angle $\varphi$ ($x_j'\rightarrow x''_j$) in order
to align the $z''$ axis with the vertical direction
\begin{equation}
\varphi=\tan^{-1}\left({\lambda_z\over\lambda_x}\tan\theta\right)\;.
\label{eq:rotation}
\end{equation}

\end{itemize}

The transformed potential reads (to simplify the notations 
in the following we omit the apices, $x_j''\rightarrow x_j$)
\begin{equation}
U_i({\bf x})={1\over2}m_i\omega_i^2\left(r^2+(z-z_{0i})^2\right)+U_{0i}
\end{equation}
where we have defined $r^2=x^2+y^2$, and 
\begin{equation}
U_{0i}=\mu_B g_{Fi} m_{Fi} B_0-{1\over2}m_i{g^2l^2\over\omega_i^2}
\end{equation}
\begin{equation}
z_{0i}=-{g l\over\omega_i^2}
\end{equation}
where the scaling factor $l$ is given by
\begin{equation}
l={\cos\theta\over\lambda_z\cos\varphi}
\end{equation}

Then we perform a translation along $z$ 
by $z_{01}$, defining $dz=z_{02}-z_{01}$,
 and we express all quantities in dimensionless units,
rescaling lengths by $a_{ho}\equiv\sqrt{\hbar/(m\omega_1)}$ and 
energies by $\hbar\omega_1$. Finally, the expressions for the
trapping potential that will be used in the rest of the paper are
\begin{eqnarray}
V_1({\bf x})\equiv U_1({\bf x})-U_{01}&=&{1\over2}\left(r^2+z^2\right)\\
V_2({\bf x})\equiv U_2({\bf x})-U_{02}&=&{1\over2}\eta\left(r^2+(z-dz)^2\right)
\end{eqnarray}
with
\begin{eqnarray}
\eta&=&{m_2\omega_2^2\over m_1\omega_1^2}
={g_{F2}m_{F2}\over g_{F1}m_{F1}}\\
dz&=&{lg\over a_{ho}}
\left({1\over\omega_2^2}-{1\over\omega_1^2}\right)=
{lg\over a_{ho}\omega_1^2}
\left({m_2\over\eta m_1}-1\right).
\end{eqnarray}

To summarize, in this section we have shown that with suitable 
transformations the trapping potential for the two condensates 
can be reduced to a simple spherical form. This allows for a much 
easier investigation of the features of the interacting system, 
as will be discussed in the following section.

\section{Thomas-Fermi approximation}
\label{sec:tf}

For large number of atoms $N_i$ the solution of Eqs. 
(\ref{eq:gpe1})-(\ref{eq:gpe2}) can be derived in the so called
Thomas-Fermi approximation which amounts to neglecting the
kinetic terms $\nabla^2\psi_i$. 
Therefore, by reabsorbing the values $U_{0i}$ of the potentials on their minima
in the definition of the chemical potentials, $\mu_i-U_{0i}\rightarrow\mu_i$,
the above equations become
\begin{eqnarray}
V_1({\bf x})+u_{11}|\psi_1|^2+u_{12}|\psi_2|^2&=&\mu_1
\label{eq:tf1}\\
V_2({\bf x})+u_{21}|\psi_1|^2+u_{22}|\psi_2|^2&=&\mu_2
\label{eq:tf2}
\end{eqnarray}
where the reduced coupling constants $u_{ij}$ are
\begin{eqnarray}
u_{11}&=&4\pi{a_{11}\over a_{ho}}\\
u_{12}&=&2\pi{a_{12}\over a_{ho}}\left(1+{m_1\over m_2}\right)=u_{21}\\
u_{22}&=&4\pi{a_{22}\over a_{ho}}{m_1\over m_2}
\end{eqnarray}

By defining $\gamma_1\equiv u_{21}/u_{11}$, 
$\gamma_2\equiv u_{12}/u_{22}$ and
$\Delta= u_{11}u_{22}- u_{12}^2$,
the solution of Eqs. (\ref{eq:tf1})-(\ref{eq:tf2}) in the overlapping 
region take the form
\begin{eqnarray}	
|\psi_1|^2&=&\alpha_{1}
\left(R_1^2 -r^2-(z-z_{c1})^2\right)
\label{eq:over1}\\
|\psi_2|^2&=&\alpha_{2}
\left(R_2^2 -r^2-(z-z_{c2})^2\right)
\label{eq:over2}
\end{eqnarray}	
where we have defined the radii $R_i$
\begin{eqnarray}	
R_1^2(\mu_1,\mu_2)&=&{2(\mu_1-\gamma_2\mu_2)\over1-\eta\gamma_2}
+{\eta\gamma_2\over(1-\eta\gamma_2)^2}dz^2\\
R_2^2(\mu_1,\mu_2)&=&{2(\mu_2-\gamma_1\mu_1)\over\eta-\gamma_1}
+{\eta\gamma_1\over(\eta-\gamma_1)^2}dz^2
\end{eqnarray}	
the position of the centers along $z$
\begin{eqnarray}	
z_{c1}&=&{-\eta\gamma_2\over1-\eta\gamma_2}dz\\
z_{c2}&=&{\eta\over\eta-\gamma_1}dz
\end{eqnarray}	
and the normalization factors $\alpha_i$
\begin{eqnarray}	
\alpha_{1}&=&u_{22}{1-\eta\gamma_2\over2\Delta}\\
\alpha_{2}&=&u_{11}{\eta-\gamma_1\over2\Delta}.
\end{eqnarray}	
Notice that in order to have overlap between $\psi_1$ and $\psi_2$
Eqs. (\ref{eq:over1})-(\ref{eq:over2}) have both to be satisfied,
that is both right members must be positive
($|\psi_1|^2,|\psi_2|^2\ge0$). The overlapping region
between the two condensates 
is therefore the intersection of the regions of space delimited
by the spherical surfaces $\Sigma_i$ defined by the equation
$R_i^2=r^2+(z-z_{ci})^2$, and identified by the sign
of the coefficient $\alpha_i$: for $\alpha_i>0$ the region
to be considered is the one inside the surface $\Sigma_i$,
for $\alpha_i<0$ the one outside.

In the regions where there is not overlap the wave functions
take the usual form
\begin{eqnarray}	
|\psi_{01}|^2&=&{1\over2u_{11}}
\left(2\mu_1 -r^2-z^2\right)\\
|\psi_{02}|^2&=&{\eta\over2u_{22}}
\left({\mu_2\over\eta} -r^2-(z-dz)^2\right)\;.
\end{eqnarray}	
Analogously to the overlapping case, these solutions are defined in a 
region of space whose boundary is delimited by the surfaces $\Sigma_{0i}$
of equation $R_{0i}^2=r^2+(z-z^0_{ci})^2$, with $R_{01}^2=2\mu_1$,
$R_{02}^2=2\mu_2/\eta$,
$z^0_{c1}=0$ and $z^0_{c2}=dz$.

Notice that in order to satisfy the continuity condition of the wave function
$\psi_i$ at the interface between the overlapping and non-overlapping
regions the wave function $\psi_{01}$ must be connected to $\psi_1$
at the boundary defined by $\Sigma_{2}$ (where $|\psi_{2}|^2$ vanishes,
but not $|\psi_{1}|^2$), and vice-versa.

\subsection{General considerations}
\label{sec:general}

\begin{figure}[htb]
\centerline{\includegraphics[height=7.5cm,clip=,angle=-90]{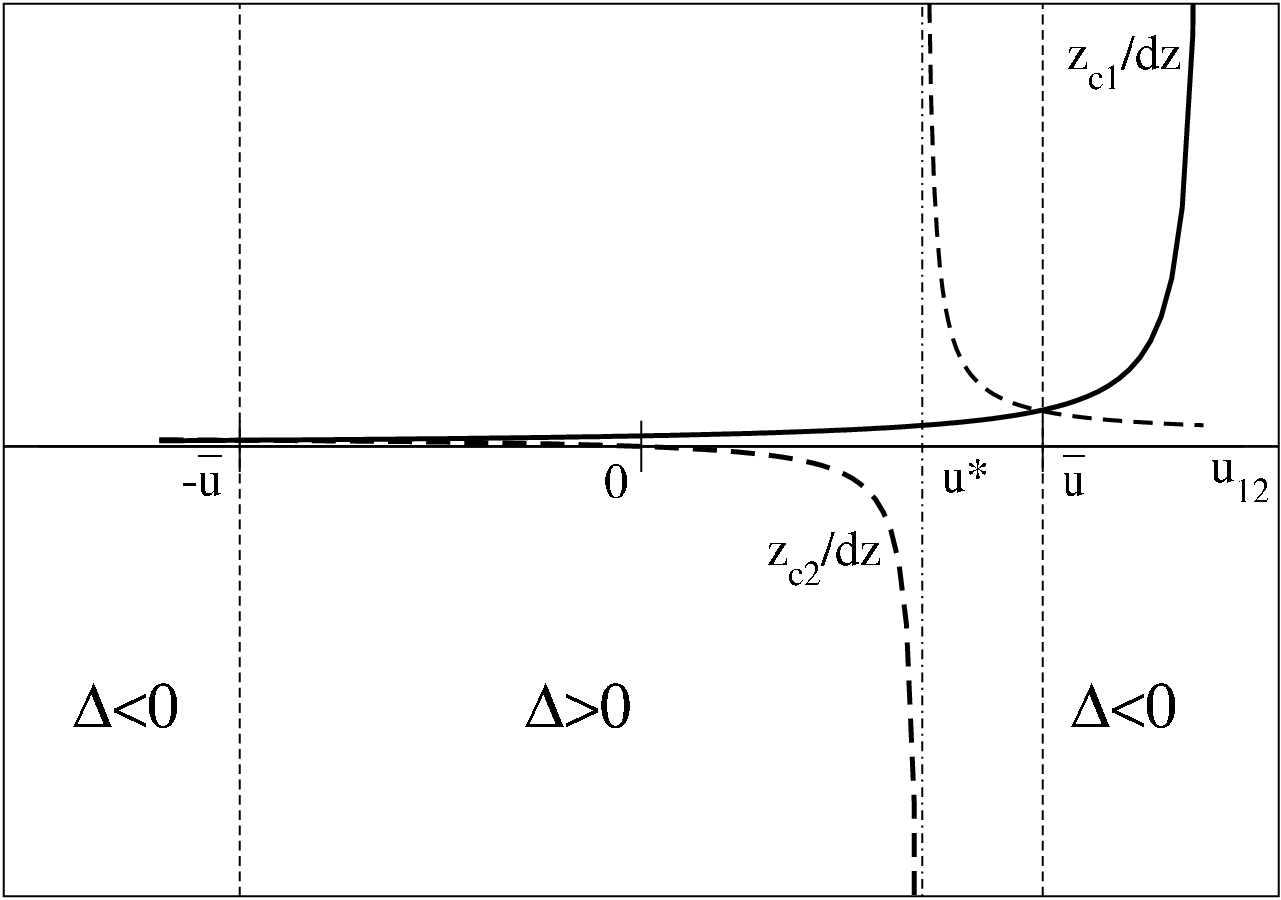}} 
\caption{Plot of the rescaled position $z_{c1}/dz$ (continuous line)   and 
$z_{c2}/dz$ (dashed line) of the centers of
the ``interacting'' surfaces $\Sigma_i$ as a function of the mutual 
coupling $u_{12}$.} 
\label{fig:centri}
\vspace{1cm}
\centerline{\includegraphics[height=7.5cm,clip=,angle=-90]{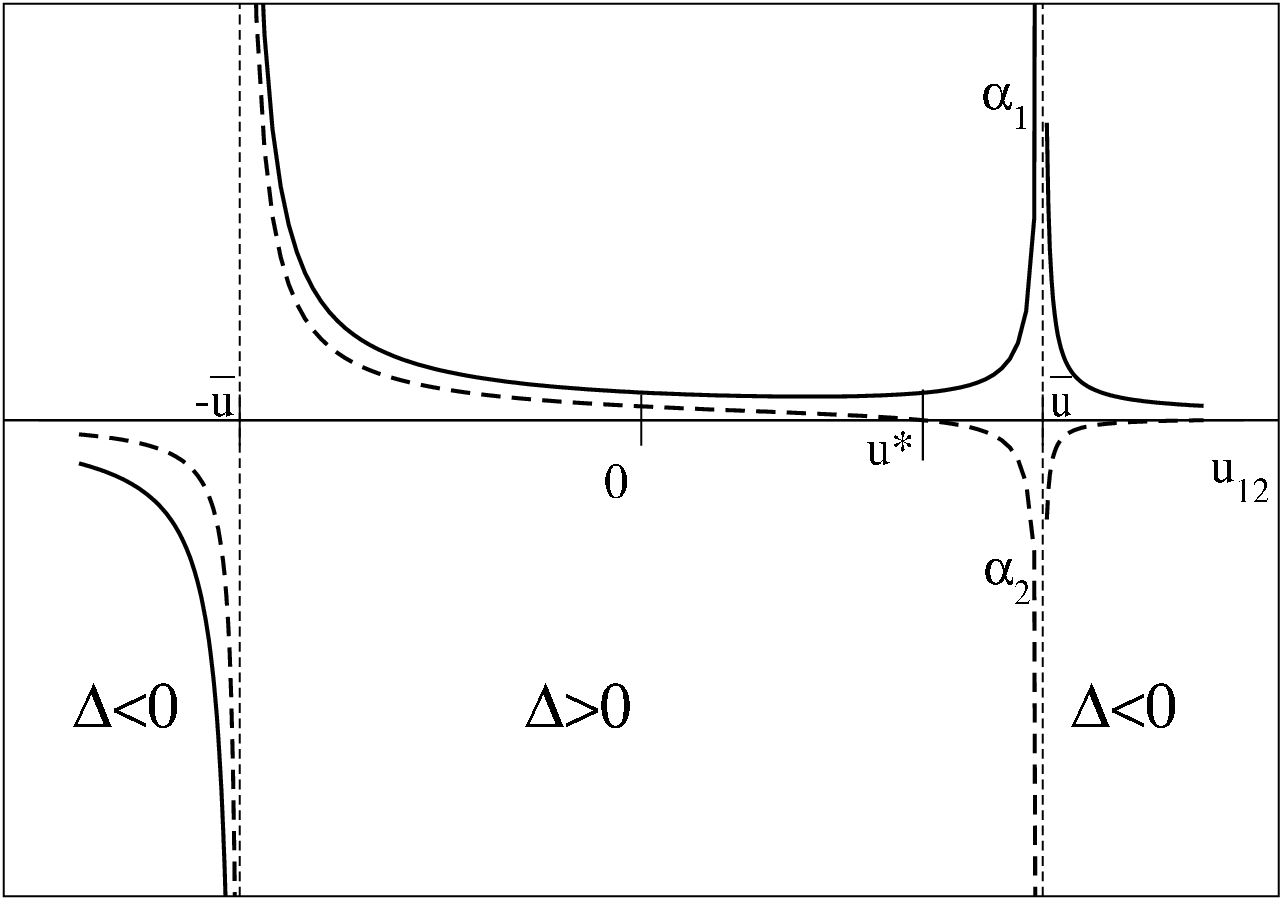}} 
\caption{Plot of the normalization factors $\alpha_{1}$ (continuous line) 
and $\alpha_{2}$ (dashed line) of
the interacting wave functions
as a function of $u_{12}$ (in arbitrary
units).} 
\label{fig:alpha}
\end{figure}

Even though a self consistent solution of the full problem
can be obtained only after having imposed the normalization of
the wave functions, we can draw some general considerations by
considering the role played by the determinant $\Delta$ and 
the coupling $u_{12}$.
First of all we define the value of $u_{12}$ at which the 
determinant $\Delta$ change sign, $\bar{u}\equiv\sqrt{u_{11}u_{22}}$.
Then we notice that the behavior of the position of the centers 
along $z$, $z_{ci}$, and the normalization factors 
$\alpha_i$ of the overlapping wave functions depends on two
critical values  $u_{12}=\eta u_{11}$ and $u_{12}=u_{22}/\eta$,
which define the poles
of $z_{ci}$ and the zeros of $\alpha_i$ (the latter have poles also for
$u_{12}=\pm\bar{u}$.
It is not difficult to prove that one of this two values lies 
in the interval where $\Delta>0$, and the other outside. 
Therefore, to fix the hierarchy of the scattering lengths we
choose the condensate 1 in order to satisfy the condition
$\eta^2\ge u_{11}/u_{22}$; with this choice
the critical value lying in the interval of positive $\Delta$
is $u^*\equiv \eta u_{11}$.
In Figs. \ref{fig:centri} and  \ref{fig:alpha} we show the position 
of the centers $z_{ci}$ and the normalization factors 
$\alpha_i$ as a function of $u_{12}$.
Notice that in correspondence of $u^*$ the center $z_{c2}$ of the 
surface $\Sigma_2$ goes from $-\infty$ to $+\infty$, and the 
normalization factor $\alpha_2$ becomes negative. Therefore for 
$u^*<u_{12}<\bar{u}$ the region where $|\psi_2|^2>0$ is the one
outside the surface $\Sigma_2$ (see also Figs. \ref{fig:topology}c,g). 

We can distinguish three cases: 
\begin{itemize}
\item[(i)]
$u_{12}<-\bar{u}$, $\Delta<0$: no overlapping solution is 
allowed in this range. From Fig. \ref{fig:alpha} we see that both
$\alpha_i$ are negative, and  therefore it is not difficult to 
prove that an overlapping region could be constructed only at the price
of putting an hole in the condensate, where both  $\psi_{i}$ would be
 vanishing.
This has obviously no physical meaning and in fact what actually happens 
is that when $u_{12}$ approaches $-\bar{u}^+$ the condensates eventually 
collapse \cite{bigelow,esry}.

\item[(ii)]
$-\bar{u}<u_{12}<\bar{u}$, $\Delta>0$: in this range
the two condensate can coexist and overlap in some region of
space if $|dz|<R_{10}+R_{20}$. We will discuss in detail the actual 
degree of overlap
and its topology in the next section.

\item[(iii)]
$u_{12}>\bar{u}$, $\Delta<0$: in this case the strong mutual repulsion
leads to a phase separation between the two condensates \cite{timm,ao}. 
The actual shape
of the interface is determined by the one at the critical value $\bar{u}$.
Since for this value the overlap goes to zero (in the TF approximation), 
if one further increases 
$u_{12}$ the shape of the interface cannot change.

\end{itemize}

Of course, if one retains the kinetic term in the GP equations
this can in part affect the degree of overlap between the two condensates.
In particular the transition
to the phase separation regime is not so sharp: the condensates can have 
an appreciable overlap also for $u_{12}\agt\bar{u}$ 
\cite{tripp,barankov,esry}. The effect of the kinetic energy is also
to raise the critical value below which the system collapses.

\subsection{Topology of spatial configurations}
\label{sec:top}

\begin{figure}[htb]
\centerline{\includegraphics[width=8cm,clip=]{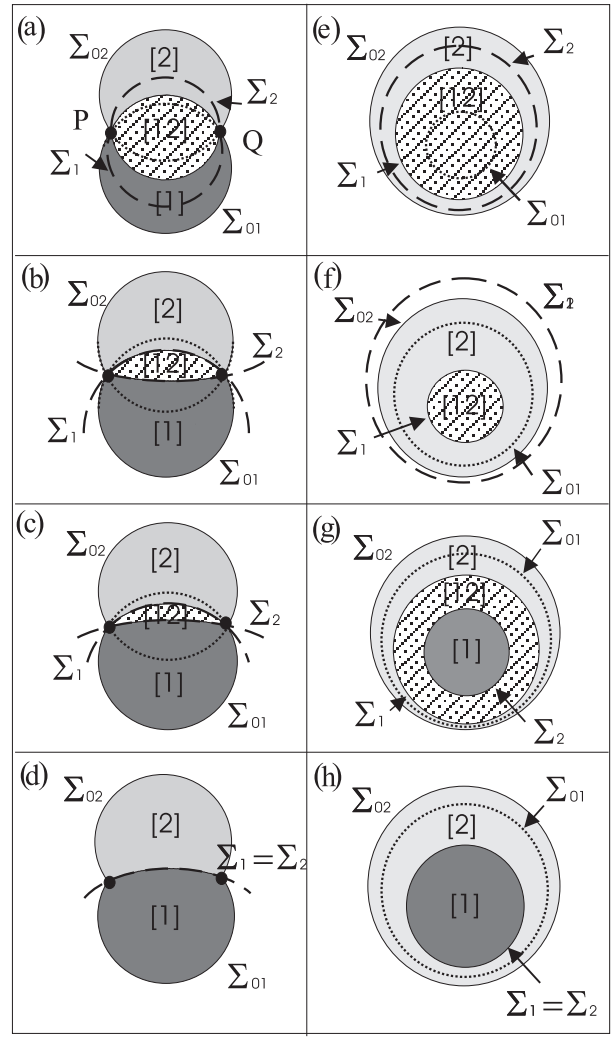}} 
\caption{
Possible topologies for a binary mixture of two BECs.
(1) ``external overlap'':  $u_{12}<0$ (a), $0<u_{12}<u^*$ (b),
 $u^*<u_{12}<\bar{u}$ (c), and phase separation $u_{12}=\bar{u}$ (d);
(2) ``full overlap'': $u_{12}<0$ (e), $0<u_{12}<u^*$ (f);
(3) ``partial overlap'': $u^*<u_{12}<\bar{u}$ (g) and phase separation 
$u_{12}=\bar{u}$ (h).
Dark and light grey represent the regions occupied by the
non-interacting condensates 1 and 2 respectively. The shaded area
indicates the overlapping region. The boundary of these regions
are delimited by the surfaces $\Sigma_{0i}$ (non-interacting, 
continuous and dotted lines)
 and $\Sigma_{i}$ (overlapping, dashed lines).
}
\label{fig:topology}
\end{figure}

In this section we investigate the different configurations which can be
obtained in the case (ii) discussed above
($\Delta>0$, $-\bar{u}<u_{12}<\bar{u}$).
Before solving completely the system for some particular set
of parameters, we give an overview of the different topologies that one
can obtain. We again distinguish three cases, as shown in 
Fig. \ref{fig:topology}

\begin{itemize}
\item[(1)]
``external overlap'':
This case can take place when 
the separation $|dz|$ between the centers is larger than the difference 
of the radii of the non-interacting profiles,
$|R_{01}-R_{02}|<|dz|<R_{01}+R_{02}$ (see Figs. \ref{fig:topology}a-d).
One can easily verify that all the four surfaces $\Sigma_i$ and 
$\Sigma_{0i}$ intersect on a circle perpendicular to the plane in 
Fig. \ref{fig:topology}, passing for the points P and Q 
(shown as black dots). 
The overlapping region is the
one contained between the surfaces $\Sigma_1$ and $\Sigma_2$
(dashed lines in the figure) whose actual shape depend on $u_{12}$,
as shown in Fig. \ref{fig:topology} for $u_{12}<0$ (a), $0<u_{12}<u^*$  (b),
 $u^*<u_{12}<\bar{u}$ (c), and $u_{12}=\bar{u}$ which is a case of phase
separation (d). For smaller $dz$ one obtains other configurations, 
which fall in the next two classes.

\item[(2)]
``full overlap'': in this case,
for $|dz|<|R_{01}-R_{02}|$, one of the two condensates is entirely
contained into the other with whom it is fully overlapping.
See Figs. \ref{fig:topology}e-f for $u_{12}<0$ and $0<u_{12}<u^*$
respectively. 
Which of the two condensates lies in the outer shell depends the actual
value of the parameters \cite{ho}.

\item[(3)]
``partial overlap'': this is similar to the case (2), but now
the overlap take place over a shell which separates the inner core
containing the condensate 1, and the outer region with the condensate 2
(Fig. \ref{fig:topology}g).
Notice that according to the above discussion this configuration 
is possible only for $u^*<u_{12}<\bar{u}$ where the sign of $\alpha_2$
is negative. This is a necessary but not sufficient condition, since also
the condition $R_{2}<R_{1}$ must be satisfied. In this range of 
$u_{12}$ (where $\alpha_2<0$) another possible solution is $R_{2}^2<0$, 
which leads to the case (2).
By further increasing $u_{12}$ to the critical value $\bar{u}$ one again 
obtain a configuration of phase separation (Fig. \ref{fig:topology}h).

\end{itemize}

Having determined the possible configurations of the system, we are now ready
to solve any particular problem by imposing the normalization condition 
(\ref{eq:norma}). To do this one has to write the normalization integrals
for each of the possible profiles in 
Fig. \ref{fig:topology}, and then solve 
Eq. (\ref{eq:norma}) in order to find the chemical potentials $\mu_i$
as a function of the atom numbers $N_i$. The analytic 
expressions for these integrals are given in Appendix \ref{app:integral}. 
These are polynomial functions of fractional powers in the chemical potentials
$\mu_i$, and in general Eq. (\ref{eq:norma}) does not admit analytical 
solutions. Therefore the relation between $\mu_i$ and $N_i$ must be inverted 
numerically (which is nevertheless a much easier task than solving
numerically the full Gross-Pitaevskii problem).
For the special case of phase separation the two normalization
equations can be decoupled (by using the fact that $R_1=R_2$ and 
$z_{c1}=z_{c2}$ for 
$u_{12}=\bar{u}$), and solved analytically for $dz=0$.

Notice that in general 
(except for some particular case, {\em e.g.} $dz=0$)
it is not possible to know {\em a priori} which of 
the various configurations in Fig. \ref{fig:topology}
applies: one has to solve Eq. (\ref{eq:norma}) for all the possible
configurations, and then choose the one which gives a self-consistent
solution. 

In summary the ground state configuration  for a particular system can
be found in three steps:
\begin{itemize}
\item[i)]
choose the normalization integrals
which applies to the possible profiles in 
Fig. \ref{fig:topology} for a given $u_{12}$, and 
determine  $\mu_i(N_i)$ by solving Eq. (\ref{eq:norma})
self-consistently;

\item[ii)]
identify the overlapping region by plotting
the ``interacting'' surfaces $\Sigma_i$;

\item[iii)]
determine the non-interacting region for each condensate
by using the ``non-interacting'' surfaces $\Sigma_{0i}$,
and the continuity of the wave functions (see Fig. \ref{fig:topology}).

\end{itemize}

We also remind that for the very special case $dz=0$ there are
also symmetry breaking solutions, not included in the present
analysis, which could be energetically favorable \cite{tripp}.

\subsection{Examples}
\label{sec:examples}

To give some explicit example we now consider two condensates of
$^{87}$Rb and $^{41}$K, 
which is a promising system for the 
realization of a new binary mixture of BECs \cite{potassio,potassio_new}. 
We will classify some possible configurations
which can be obtained by varying the number of atoms in each 
condensate, for different values of the inter-atomic scattering
length, which is considered here as a tunable parameter \cite{a12}. 
The results, valid in the Thomas-Fermi (TF) approximation,
will be compared with the numeric solution of the full-3D Gross-Pitaevskii
equations (GPE), found using a steepest descent method \cite{report}.

\begin{figure}[htb]
\centerline{\includegraphics[width=8cm,clip=,bb=230 620 500 720]{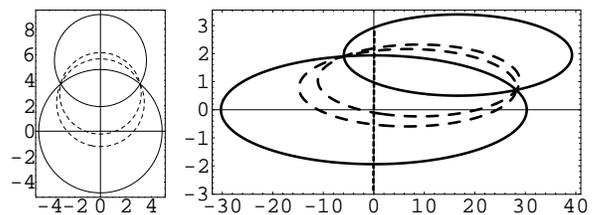}}
\caption{TF profiles of
the two condensates (non-interacting: continuous; overlapping: dashed)
in rescaled
(left) and natural (right, in units of $a_{ho}$) coordinates
($x$ horizontal, $z$ vertical)
for a case of ``external overlap'':
$N_{Rb}=5\cdot10^{4}$, $N_{K}=2\cdot10^{4}$,
$a_{12}=-55~a_0$.}
\label{fig:rotation}
\end{figure}

We start by considering a case of ``external overlap'' with
both condensates in the hyperfine level $|2,2\rangle$ ($\eta=1$).
The scattering lengths are $a_{Rb}=99~a_0$ and $a_{K}=60~a_0$, 
$a_0$ being the Bohr radius \cite{scatt}. 
As trap frequencies we use 
$\omega_x^{Rb}=16~$Hz, $\omega_y^{Rb}=\omega_z^{Rb}=250~$Hz,
with an angle of rotation
$\theta=0.035$ (we retain these values for all the cases
analyzed in this section).
With this choice the reduced coupling constant $u_{ii}$ are
\begin{equation}
u_{Rb,Rb}=0.0611\;,\quad
u_{K,K}=0.0785
\end{equation}
and therefore, according to the above discussion, we identify
the condensates 1 with $^{87}$Rb, and  the condensates 2 with $^{41}$K.
We choose a case attractive interaction between the two condensates, 
$a_{12}=-55~a_0$ ($u_{12}=-0.0530$),
with  $N_{Rb}=5\cdot10^{4}$ and $N_{K}=2\cdot10^{4}$. 
To visualize the role of the scaling and rotation transformations,
in Fig. \ref{fig:rotation} we show the TF profiles of
the two condensates in the $x-z$ plane, in rescaled (left) and 
natural  coordinates (the coordinate axes 
correspond to the trap eigenaxes; right).
The profiles in natural units can be easily obtained by performing 
the inverse  transformation of those in Eqs. (\ref{eq:scaling})  
and (\ref{eq:rotation}). We will use this system of coordinates
to show all the following figures.
Notice that despite the small rotation angle $\theta$,
(the direction of gravity, represented by a dotted line in
the right picture of Fig. \ref{fig:rotation}, 
is almost indistinguishable from the $z$ axis on the scale of
the figure) the misalignment in the 
direction of gravity produces a relatively large horizontal sag in the $x$
direction where the trap confinement is weak.

\begin{figure}[htb]
\centerline{\includegraphics[width=8cm,clip=,bb=100 360 430 715]{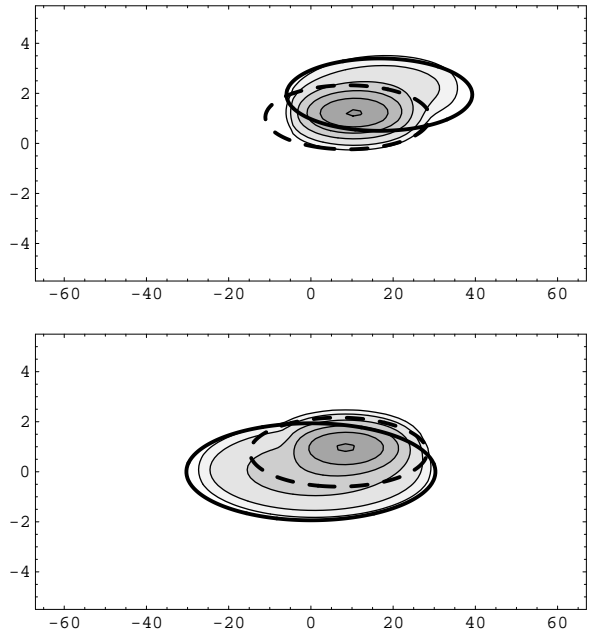}}
\caption{Density contours of the GPE solutions in the 
$x-z$ plane for the 
$^{87}$Rb (bottom) and $^{41}$K (top) condensates. 
Each condensate is compared with 
the TF profiles which define the boundary of the non-interacting 
or overlapping phases,
as defined in Fig. \ref{fig:topology}a.
This a case of ``external overlap'' with attractive interaction 
between the two condensates, $a_{12}=-55~a_0$, 
and  $N_{Rb}=5\cdot10^{4}$, $N_{K}=2\cdot10^{4}$. Lengths are given 
in units of $a_{ho}$.  }
\label{fig:ex-ext}
\end{figure}

In Fig. \ref{fig:ex-ext} we compare the TF profiles with the 
contour plot of the two densities, as found from the full GPE solution.
For clarity each condensate is plotted separately, and compared with 
the TF profiles which define the boundary of the non-interacting 
(continuous lines) or overlapping phases (dashed lines),
as defined in Fig. \ref{fig:topology}.
The outer contour line for each condensate correspond 
to $10\%$ of its peak density (for $y=0$).

\begin{figure}[htb]
\centerline{\includegraphics[width=7.5cm,clip=,bb=100 500 345 770]{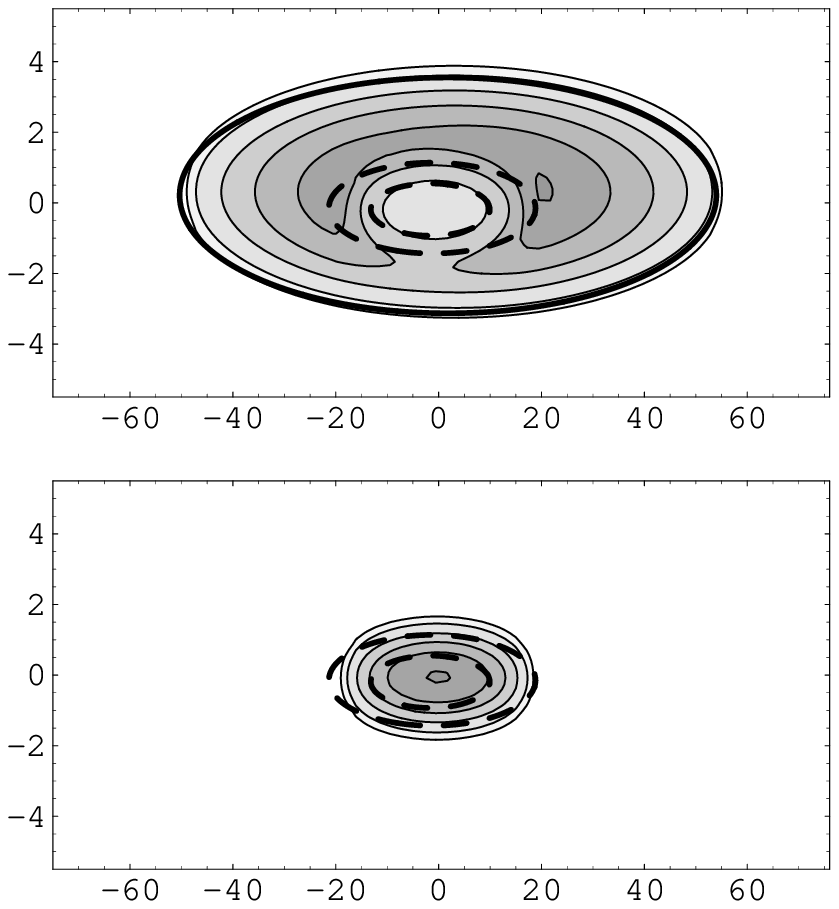}}
\caption{Density contours of the GPE solutions in the 
$x-z$ plane for the
$^{87}$Rb (bottom) and $^{41}$K (top) condensates,
for a case of ``partial overlap'': $N_{Rb}=2\cdot10^{4}$, 
$N_{K}=2\cdot10^{5}$,
$a_{12}=67~a_0$. Each condensate is compared with 
the TF profiles which define the boundary of the non-interacting or 
overlapping phases,
as defined in Fig. \ref{fig:topology}g.
Lengths are given in units of $a_{ho}$.}
\label{fig:ex-part}
\end{figure}

Then we consider two examples for a system where the $^{87}$Rb condensate
 is in the hyperfine level $|2,2\rangle$ 
and the $^{41}$K condensate in $|2,1\rangle$
(we use again $a_{K,K}=60~a_0$). In this case $\eta=0.5$.
In Fig. \ref{fig:ex-part} we show a case 
of ``partial overlap'', obtained by fixing the inter-atomic 
scattering length  to 
$a_{12}=67~a_0$ ($u_{12}=0.0645$)
and the number of atoms to  $N_{Rb}=2\cdot10^{4}$ and
$N_{K}=2\cdot10^{5}$. Notice that when both condensates
are in the $|2,2\rangle$ level the spatial separation
between the two is too large to allow for a configuration
of ``partial overlap'' for reasonable values of the trap frequencies
(in principle one could reduce the separation by strongly increasing
the confinement in the direction of gravity).

\begin{figure}[htb]
\centerline{\includegraphics[width=8cm,clip=,bb=100 370 420 720]{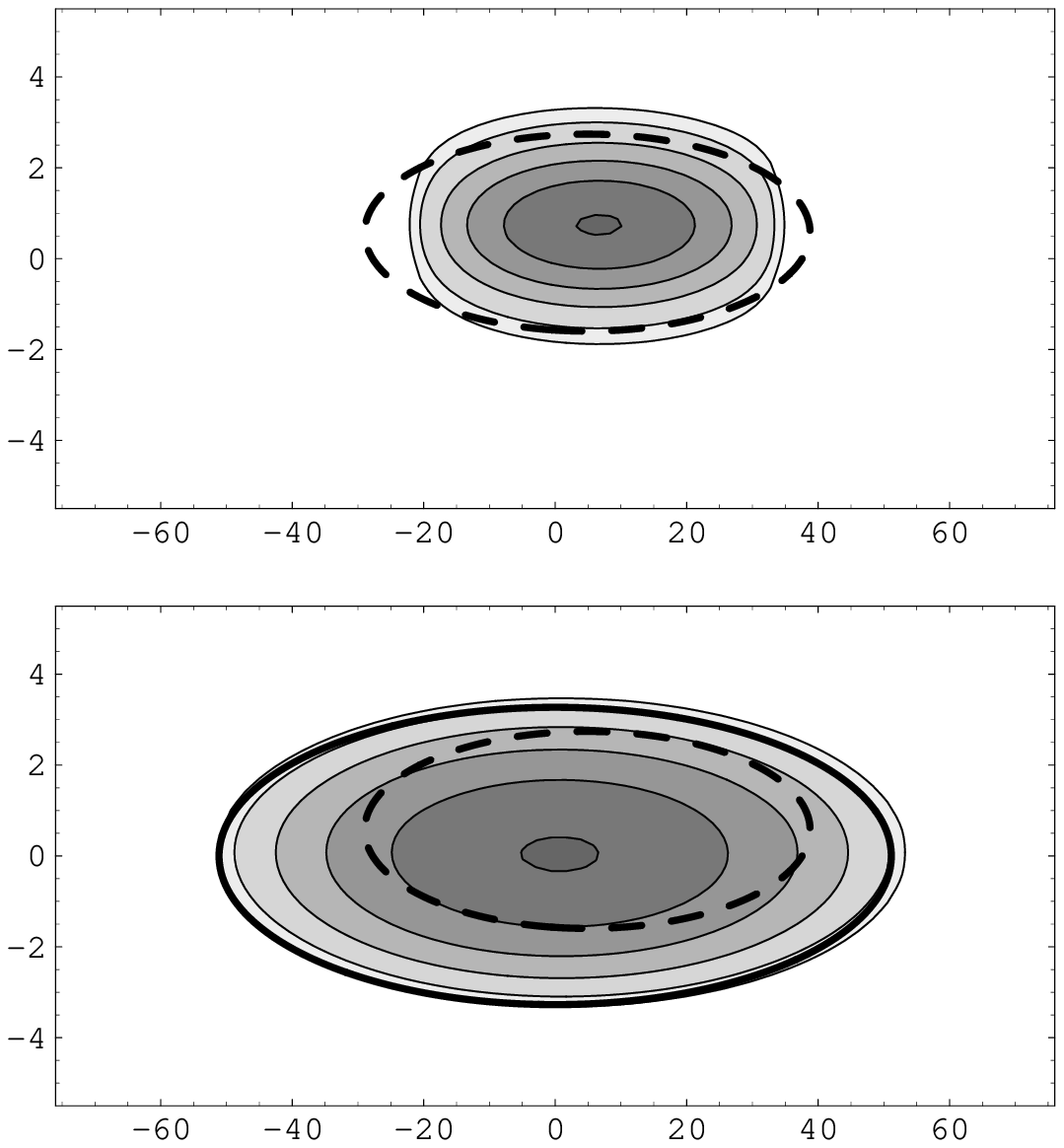}}
\caption{Density contours of the GPE solutions in the 
$x-z$ plane for the $^{87}$Rb (bottom) and $^{41}$K (top) condensates,
for a case of ``full overlap'': $N_{Rb}=5\cdot10^{5}$, $N_{K}=1\cdot10^{4}$,
$a_{12}=20~a_0$. Each condensate is compared with 
the TF profiles which define the boundary of the non-interacting 
or overlapping phases,
as defined in Fig. \ref{fig:topology}f.
Lengths are given in units of $a_{ho}$.
}
\label{fig:ex-full}
\end{figure}

Finally, in Fig. \ref{fig:ex-full} we show
a case of ``full overlap'' for  $N_{Rb}=5\cdot10^{5}$, 
$N_{K}=1\cdot10^{4}$, and
$a_{12}=20~a_0$, giving $u_{12}=0.0193$. 
 
From the examples considered here we see that, although the full 
solution of the GPE is required for a precise determination of
the actual degree of overlap between the two condensates, the
TF approximation well captures the basic topology of the
ground state configurations. Therefor, due to its simplicity,
the TF method presented here can be a useful tool to characterize
the ground state structure of a binary mixture of BECs also in
presence of a non trivial geometry.

We conclude this section by noting that we have also verified that 
our method well reproduces the results already studied in literature 
in case of simpler geometries \cite{ho,esry,tripp}.

\section{Conclusions}
\label{sec:conclusions}

We have presented a general method to classify the
ground-state of a binary mixture of Bose-Einstein condensates.
We have considered the general case of of two different atomic species,
with different mass and in different hyperfine states, 
trapped in a magnetic potential. We have explicitly included the
possibility of a non trivial gravitational sag, when the direction
of gravity is not aligned with any of the trap eigenaxes, since
even a small misalignment can produce a large ``horizontal sag''.
We have shown that, within the Thomas-Fermi approximations
and by performing a suitable coordinate transformation, the 
equations for the density distributions can be put in a simple 
spherical form. 
We have given explicit rules to classify
the different spatial topologies which can be produced, 
and we have discussed how the behavior of the system is influenced by the 
inter-atomic interaction.
 
We have also provided explicit examples, and compared the results with
the full numeric Gross-Pitaevskii calculation, finding a good agreement.

The results presented in this paper might be useful for analyzing 
future experiments where new combinations of 
binary condensates are likely to be produced \cite{potassio,potassio_new}.

\begin{acknowledgments}
We acknowledge useful discussion with G. Modugno and G. Roati.
\end{acknowledgments}

\vspace{2cm}
{\it Note added} After having completed this work we became 
aware of a very recent preprint related to this subject \cite{jezek}.
\appendix

\section{Normalization integrals}
\label{app:integral}

In this appendix we give the explicit expressions for the integrals
which enter the normalization condition (\ref{eq:norma}). We
distinguish two general cases: (i) ``internal overlap'' one of the 
two condensates is entirely contained into the other 
(Fig. \ref{fig:topology}, right column), and (ii) ``external overlap'',
(Fig. \ref{fig:topology}, left column).

In both cases the normalization condition can be 
written as a sum of integrals of a generic density
\begin{equation}
|\psi|^2=\alpha\left(R_c^2 -r^2-(z-z_c)^2\right)
\end{equation}
over an appropriate portions of spherical domain whose boundary is 
given by the surface
\begin{equation}
R_A^2 = r^2+(z-z_A)^2\;.
\end{equation}
In the following subsections we consider explicitly the two cases.

\subsection{Internal overlap}
\label{app:internal}

In this case the normalization condition can be imposed by using a 
combination of integrals over {\em spherical} domains.  The generic form is
\begin{eqnarray}
&&I_I(\alpha,z_c,R_c,z_A,R_A)=
\label{eq:master-int}\\
&&\quad4\pi\alpha R_A^3\left[{1\over3}R_c^2-{1\over3}(z_c-z_A)^2 
-{1\over5}R_A^2\right]\nonumber
\end{eqnarray}
From this expression one also recover the value of the integral for the
non-interacting case
\begin{equation}
I_n(\mu_i)={8\pi\over15}\alpha R_c^5
\end{equation}
($I_n(\mu_i)=4\pi(2\mu_i)^{5/2}/u_{ii}$ for $\eta=1$).

By using appropriate combinations of the integral (\ref{eq:master-int}), 
the normalization condition for the case shown in 
Fig. \ref{fig:topology}g reads
\begin{widetext}
\begin{eqnarray}
N_1&=&
I_I(\alpha_{1},z_{c1},R_1(\mu_1,\mu_2),z_{c1},R_1(\mu_1,\mu_2))
-I_I(\alpha_{1},z_{c1},R_1(\mu_1,\mu_2),z_{c2},R_2(\mu_1,\mu_2))\nonumber\\
&&
+I_I(\alpha_{01},0,\sqrt{2\mu_1/\eta},z_{c2},R_2(\mu_1,\mu_2))\\
N_2&=&I_n(\mu_2)
- I_I(\alpha_{02},dz,\sqrt{2\mu_2/\eta},z_{c1},R_1(\mu_1,\mu_2)\nonumber\\
&&+I_I(\alpha_{2},z_{c2},R_2(\mu_1,\mu_2),z_{c1},R_1(\mu_1,\mu_2))
-I_I(\alpha_{2},z_{c2},R_2(\mu_1,\mu_2),z_{c2},R_2(\mu_1,\mu_2))
\end{eqnarray}
where we have indicated the explicit dependence on $\mu_1$ and $\mu_2$.
The cases in Figs. \ref{fig:topology}e,f,h can be constructed 
in a similar way.

\subsection{External overlap}
\label{app:external}

These are the configurations shown in Fig. \ref{fig:topology}a-d. In this
case the master integral can be written as the integral over a {\em convex} 
domain delimited by two spherical surfaces 
(A and B the upper and lower ones along the $z$ axis respectively)
\begin{eqnarray}
&&I_E(\alpha,z_c,R_c,z_A,R_A,z_B,R_B)=
\pi\alpha\left[R_A^2\left(R_c^2-0.5R_A^2-(z_c-z_A)^2\right)
(R_A-\bar{z}(A,B))\right.\nonumber\\
&&\quad\left.-R_A^2(z_c-z_A)(R_A^2-\bar{z}^2(A,B))
-{1\over3}\left(R_c^2-(z_c-z_A)^2\right)(R_A^3-\bar{z}^3(A,B))
\right.\nonumber\\
&&\quad\left.+{1\over2}(z_c-z_A)(R_A^4-\bar{z}^4(A,B))
+{1\over10}(R_A^5-\bar{z}^5(A,B))
\right] + \left(z_A\leftrightarrow z_B,R_A\leftrightarrow -R_B\right)
\label{eq:master-ext}
\end{eqnarray}
with
\begin{equation}
\bar{z}(A,B)={R_A^2-R_B^2+(z_B-z_A)^2\over2(z_A-z_B)}\;.
\end{equation}

By assuming a configuration where the condensate
1 has a lower position along $z$ (as in Fig. \ref{fig:topology}),
the normalization condition for the cases with $u_{12}<u^*$ 
shown in Fig. \ref{fig:topology}a,b is 
\begin{eqnarray}
N_1&=&I_n(\mu_1)-I_E(\alpha_{01},0,\sqrt{2\mu_1},0,\sqrt{2\mu_1},
z_{c2},R_2(\mu_1,\mu_2))\nonumber\\
&&+I_E(\alpha_{1},z_{c1},R_1(\mu_1,\mu_2),
z_{c1},R_1(\mu_1,\mu_2),z_{c2},R_2(\mu_1,\mu_2))\\
N_2&=&I_n(\mu_2)-I_E(\alpha_{02},dz,\sqrt{2\mu_2/\eta},
z_{c1},R_1(\mu_1,\mu_2),dz,\sqrt{2\mu_1})\nonumber\\
&&+I_E(\alpha_{2},z_{c2},R_2(\mu_1,\mu_2),
z_{c1},R_1(\mu_1,\mu_2),z_{c2},R_2(\mu_1,\mu_2))
\end{eqnarray}
\end{widetext}
In an analogous way one can construct the appropriate normalization
condition for all other cases in this class (``external overlap''),
by considering the appropriate combination of integral of the form
(\ref{eq:master-ext}) over {\em convex} domains.


\begin{thebibliography}{}
\bibitem{nota}
In general one could consider the situation in which the vertical
direction does not lie in any of symmetry planes of the trap.
Anyway, since most experiments are performed with cylindrically
symmetric traps, we restric our analysis to this case. The 
extension to the case of anisotropic traps is straightforward.
\bibitem{a12} 
Although the scattering length $a_{Rb,K}$ has been recently
estimated to be of the order of $163~a_0$ \cite{potassio_new}, 
in principle it is possible to tune its value by using 
Feshbach resonances.
\end{thebibliography}
\end{document}